\documentclass[showpacs]{revtex4}
\usepackage{fullpage}
\usepackage[dvips]{graphicx}
\input{epsf}
\usepackage{amsmath}
\usepackage{amssymb}
\usepackage[mathscr]{eucal}

\newcommand{\ket}[1]{| #1 \rangle}
\newcommand{\bra}[1]{\langle #1 |}

\begin{document}

\title{Parameter differentiation and quantum state decomposition for time varying Schr{\"{o}}dinger equations}
\author{Claudio Altafini}\thanks{This work was supported by a grant from the Foundation Blanceflor Boncompagni-Ludovisi.}
\affiliation{SISSA-ISAS  \\
International School for Advanced Studies \\
via Beirut 2-4, 34014 Trieste, Italy }
\email{altafini@sissa.it}


\begin{abstract}
For the unitary operator, solution of the Schr{\"{o}}dinger equation corresponding to a time-varying Hamiltonian, the relation between the Magnus and the product of exponentials expansions can be expressed in terms of a system of first order differential equations in the parameters of the two expansions.
A method is proposed to compute such differential equations explicitly and in a closed form.
\end{abstract}

\maketitle 


\section{Introduction}
For energy preserving finite dimensional closed quantum systems, the time evolution is well-known to be completely trivial.
In fact, the linear and hermitian hamiltonian $H$ admits the spectral decomposition $ H = \sum _{i=1}^N \epsilon_i \ket{\phi_i}\bra{\phi_i }$ with $ \epsilon_i $ the energy eigenvalues corresponding to the stationary eigenstates $ \ket{\phi_i } $, and the solution of the associated Schr\"{o}dinger equation in the $ \ket{\phi_i(t) } $ is simply obtained by adding a phase factor proportional to the corresponding energy eigenvalue: $
\ket{\phi_i (t) } = {\rm exp} \left( -i \epsilon_i t /\hbar  \right) \ket{\phi_i } $.
Even without resorting to the spectral resolution and to the change of basis that diagonalizes the Hamiltonian, the time evolution of the quantum state $ \ket{\psi} $ can always be written as $ \ket{\psi(t) } = U(t) \ket{\psi(0)} $, with the unitary propagator $
U(t) = {\rm exp} \left(-i H t /\hbar  \right) \in SU(n) $
explicitly known.
If $ A_1 , \ldots , A_n $, $n=N^2-1$, is a basis of $\mathfrak{su}(N) $, then it is possible to write $ U(t) $ in terms of some Euler-like parameterization of $SU(N) $.
These parameterizations are essentially ordered products of exponentials on $ SU(N) $ and for a Lie group there are as many such products as there are Lie group decompositions.   
See \cite{Murnaghan1,Reck1,Nemoto1} for a few examples of explicit choices on $ SU(N) $, \cite{Byrd1,Dattoli1,Mathur1,Rowe1} for $SU(3)$ and \cite{Crouch3,Lowenthal1} for the related issue of how to uniformly parameterize the rotation groups.
Such decompositions are very useful for example in the generation of elementary gates in quantum computing or more generally in the control of driven dynamics \cite{Ramakrishna3,Schirmer4}, but also in quantum state disentanglement \cite{Raghunathan1}, in the study of coherent states \cite{Mathur1,Nemoto1}, in the solution of the Liouville-von Neumann equation and in the design of schemes for the numerical integration of differential equations on Lie groups \cite{Iserles1}.

The picture becomes more complicated when the Hamiltonian $H$ is time-varying, because the energy (and thus the eigenstructure) is not conserved by the dynamics.
The two expressions for the $U(t) $ mentioned above also became time-dependent and go under the names of Magnus expansion \cite{Magnus1} and Wei-Norman expansion \cite{Wei1}.
The literature on the subject is quite vast, see \cite{Suarez1} and references therein, and deals especially with the convergence problem of the two series, which is not an issue here.

We are interested here  in studying how the two expansions relate, in particular how the parameters of the one series can be expressed as functions of the parameters of the other one.
{\em In nuce}, such a transformation is already in the original papers of Wei-Norman, but its importance is made clear in \cite{Wilcox1}.
In practice, it consists in studying a system of nonlinear differential equations having as variables the two sets of parameters contained in the two expansions.
Such a system of differential equations is called the {\em Wei-Norman formula} and it corresponds to the {\em Jacobian} of the change of coordinates i.e. of the transformation from single exponential to product of exponentials.
As such, it is linear in one of the two sets of coordinates although it depends on the point of application, expressed nonlinearly in terms of the other set of parameters.

The Wei-Norman formula appears in several different contexts in the literature, see \cite{Brockett3,Dattoli1,Huillet1,Salmistrato1} just to mention a few.
In this study, we propose a method to compute it explicitly and systematically for any dimension, based only on the structure constants of the Lie algebra.
Such method seems to be new.
The {\em rationale} behind it is a technique to compute in closed form one parameter groups of automorphisms i.e. exponentials of the matrices of the adjoint representation of any linear Lie algebra.
Such technique is adapted from \cite{Raghunathan1}.
As an example, we compute two  Wei-Norman formulae for $\mathfrak{su}(2) $ for the same basis obtained from the Pauli matrices but for different ordering of the basis elements (i.e. different choices of Euler angles) and a formula for $\mathfrak{su}(3)$.
As we will see, the explicit symbolic computations rapidly grow with the dimension $N$; however numerical simulations of the dynamics can always be carried out easily.

\section{Magnus expansion versus product of exponentials expansion}
Consider a quantum mechanical system with Hamiltonian $ H(t) $ which is a continuous function of time.
Assume $H(t)$ is Hermitian and belongs for all $t$ to the finite dimensional Lie algebra $ \mathfrak{su}(N) $.
If we choose a basis of skew-symmetric matrices $ A_1 , \ldots A_n $ for $ \mathfrak{su}(N) $, then $ [ A_i, \, A_j ] =c_{ij}^\mu A_\mu $ where $ c_{ij}^k $ are the structure constants of $ \mathfrak{su}(N) $.
In the following we will assume that the summation convention over repeated indexes is imposed only for the greek letters, not for the latin.
Since $ A_1 , \ldots, A_n $ are time independent operators in $ \mathfrak{su}(N) $, $ H(t)$ can be written in this basis as $ H(t) = u^\mu (t) A_\mu $ with $ u^i (t) $ analytic functions of time.
Let $ U(t) $ be the time evolution operator of the Schr\"{o}dinger equation:
\[
i \frac{\partial }{\partial t} \ket{\psi(t)} = H (t) \ket{\psi(t)} \qquad \ket{\psi(0)} = \ket{\psi_0 }
\]
Then $ \ket{\psi(t)} = U(t) \ket{\psi_0 }$ where $U(t) $ in the Magnus expansion is given by the formal expression
\begin{equation}
U(t) = T {\rm exp} \left( \int_{0}^t u^\mu(\tau) A_\mu  d \tau \right) 
\label{eq:Magnus-expan1}
\end{equation}
where $T$ is the Dyson operator and $ {\rm exp } $, the exponential map for $SU(N) $, is the ordinary matrix exponential.
The {\em Wei-Norman formula} relates \eqref{eq:Magnus-expan1} with the expansion as a product of exponentials, i.e. it affirms that \eqref{eq:Magnus-expan1} can be written locally around the identity of $SU(N) $ as 
\begin{equation}
U(t) = {\rm exp} \left( \gamma^1(t) A_1 \right) \ldots {\rm exp} \left( \gamma^n(t) A_n \right) 
\label{eq:WeiNorm-expan1}
\end{equation}
In other words, it is used to relate the flow of the solution of the differential equation
\[
i  \frac{\partial }{\partial t} U(t) = H (t)U(t) \qquad U(0) = I
\]
with the product of one-parameter subgroups $ \gamma^i A_i $ lifted to $ SU(N) $ and used as a coordinate chart on the group itself.
If, in the original formulation of Wei-Norman, the same set of infinitesimal generators as the basis is chosen, nobody forbids to alter the order of the $A_i $ in \eqref{eq:WeiNorm-expan1}, or even to consider any alternative fundamental representation of the group, for example one of the Euler-like parameterizations introduced in \cite{Byrd1,Murnaghan1,Mathur1,Nemoto1}.
We will stick to \eqref{eq:WeiNorm-expan1} for notational convenience, but develop the corresponding of the Wei-Norman formula for the Euler ZYZ parameterization for the $SU(2)$ case in the example of Section \ref{sec:su2}.

The Wei-Norman formula consists in expressing the functions $ \gamma^i (t) $ in terms of the $ u^i(t)$ via a system of differential equations \footnote{Notice that in the whole paper we consider only right invariant vector fields of $SU(N)$.
Thus also the Wei-Norman formula reported is right invariant.
A completely analogous left invariant version could be obtained.} 
\begin{equation}
\Xi (\gamma^1 , \ldots , \gamma^n ) \begin{bmatrix} \dot \gamma^1 \\ \vdots \\ \dot \gamma^n \end{bmatrix} =  \begin{bmatrix} u^1 \\ \vdots \\ u ^n \end{bmatrix} \qquad \gamma^i(0) = 0 
\label{eq:matrix-xi}
\end{equation}
with the $n\times n $ matrix $ \Xi $ analytic in the variables $ \gamma^i $.
Since the adjoint maps can all be written as
\begin{equation}
e^{\gamma^j {\rm ad}_{A_j} } A_i =
e^{\gamma^j A_j } A_i e^{ - \gamma^j A_j} = \sum^{\infty}_{l=0} \frac{
  (\gamma^j )^l }{l!}  {\rm ad}_{A_j}^l A_i = \sum^{\infty}_{l=0} \frac{
  (\gamma^j) ^l }{l!} c_{i \, j } ^{\mu_1} c_{i \, \mu_1} ^{\mu_2} \ldots c_{i \, \mu_{l-1} } ^{\mu_l }  A_{\mu_l} 
\label{eq:adj2}
\end{equation}
the matrix $ \Xi $ of elements $ (\Xi)_{ki}= \xi^k_i$ is defined in terms of the $ \gamma^i $ and of the structure constants as:
\begin{equation}
\prod_{j=1}^m e^{ \gamma^j {\rm ad} _{A_j } } A_i = \xi_i^\mu A_\mu  \qquad m=1, \ldots , n 
\label{eq:wei-tmp1}
\end{equation}
When the $ A_i $ form a basis of $\mathfrak{su}(N) $, the matrix $ \Xi $ assumes also the meaning of map between canonical coordinates of the first kind \eqref{eq:Magnus-expan1} and canonical coordinates of the second kind \eqref{eq:WeiNorm-expan1}, see \cite{Varadarajan1}.
In this case, since $ \gamma^i(0) =0 $, $ \Xi(0) = I $ and thus $ \Xi$ is locally invertible.
However, because of the semisimplicity of $SU(N) $, all parameterizations lead to a Wei-Norman formula that is subject to singularities and as such $ \Xi^{-1} $ has only a local validity.
By inverting $ \Xi $, equation \eqref{eq:matrix-xi} assumes the more traditional aspect of a system of first order differential equations in the $ \gamma^i $ variables:
\begin{equation}
 \begin{bmatrix} \dot \gamma^1 \\ \vdots \\ \dot \gamma^n \end{bmatrix} = \Xi (\gamma^1 , \ldots , \gamma^n )^{-1}  \begin{bmatrix} u^1 \\ \vdots \\ u ^n \end{bmatrix} \qquad \gamma^i(0) = 0 
\label{eq:matrix-xiinv}
\end{equation}
If the time evolution of one of the two vectors of coordinates $ \gamma^i $ or $ u^i $ is known, the formul\ae \eqref{eq:matrix-xi} or \eqref{eq:matrix-xiinv} can be used to obtain the other one.
While \eqref{eq:matrix-xi} is global, \eqref{eq:matrix-xiinv} is valid only as long as $ {\rm det}(\Xi)\neq 0 $ and thus the nonsingularity of $\Xi $ needs to be checked at the point of application.  
Another weak point of the Wei-Norman formula is that the system of differential equations is nonlinear in the $\gamma^i $.

Finally, \eqref{eq:adj2} as it stands involves infinite sums.
In the following paragraph we provide a method for transforming these infinite series into finite sums.

\section{Computation of the exponentials in the adjoint representation}
\label{sec:adj}

The adjoint representation of a linear Lie algebra like $ \mathfrak{su}(N)$ is a derivation of the algebra and as such it is the infinitesimal generator of a one-parameter group of automorphisms.
In the basis $ A_1, \ldots , A_n $ of the Lie algebra, the one-parameter groups of automorphisms are described by the matrix exponentials $ e^{{\rm ad}_{A_i}}$ \footnote{The infinitesimal generators of the one-parameter subgroups $ {\rm ad}_{A_1} , \ldots , {\rm ad}_{A_n} $ are often referred to as commutator superoperators \cite{Ernst1}. We will not use this terminology here.}.

There exist several techniques to compute the exponential of a square matrix.
The simplest is perhaps the use of the Sylvester formula \cite{Barnett1}, p.233, which requires only the knowledge of the eigenvalues of the matrix.
Alternatively, one can use the spectral decomposition of normal operators i.e. compute the Jordan form by similarity transformations, but this requires the knowledge of eigenvalues and eigenvectors.
The method we use here is taken from \cite{Raghunathan1}. It relies on the Cayley-Hamilton theorem and consists in expressing the series expansion of $ e^{{\rm ad}_{A_i}}$ in terms of the first $ n-1 $ powers of $ {\rm ad}_{A_i} $ with suitable coefficients depending on the coefficients of the characteristic polynomial of $ {\rm ad}_{A_i} $ and on $ \gamma_i $.
Of the different methods that can be used to express a matrix exponential, in our opinion this is the most suited for the adjoint representation, as the powers of $ {\rm ad}_{A_i} $ are immediately expressed in terms of the structure constants of the Lie algebra.
In fact, if $ c^k_{ij}$ are the structure constants of the Lie algebra,
the matrices corresponding to the basis elements $A_i $ are $ {\rm ad}_{A_i } = M_i $ of elements $ (M_i ) _{kj} = c^k_{ij} $ and the $l$-th power of $ {\rm ad}_{A_i} $ is given by 
\begin{equation}
M_i ^l= \left( M_i^l \right) _{kj} = c_{1 \, \mu_1 }^k c_{1 \, \mu_2} ^{\mu_1} \ldots c_{1 \, \mu_{l-1} }^{\mu_{l-2} } c_{1 \, j}^{\mu_{l-1}}
\label{eq:Mi-to-the-l}
\end{equation}
Any matrix $B \in \mathfrak{su}(N) $ can be written as $ B = b^\mu A_\mu $.
If we identify $B$ with the $n$-dimensional coordinate vector $ B\simeq b = \begin{bmatrix} b^1 & \ldots & b^n \end{bmatrix}^T $, then $ \left[ A_i, \, B \right] \simeq {\rm ad}_{A_i} b = (M_i ) b $, i.e. the Lie bracket gives another column vector 
\[
 (M_i) _{k\mu} b^\mu = c^k_{i\mu} b^\mu = \begin{bmatrix} c_{i  \mu }^1 b^\mu & c_{i  \mu}^2 b^\mu & \ldots & c_{i \mu }^n b^\mu \end{bmatrix}^T  
\]
while the Lie bracket of $ D \simeq d = \begin{bmatrix} d^1 & \ldots & d^n \end{bmatrix}^T $ with $B$ looks like:
\[
 \left[ D, \, B \right] \simeq {\rm ad}_D b = d^\nu (M_\nu ) b =  \begin{bmatrix}d^\nu c_{\nu  \mu }^1 b^\mu & d^\nu c_{\nu \mu}^2 b^\mu & \ldots & d^\nu c_{\nu \mu }^n b^\mu \end{bmatrix}^T
\]
As an example, compute $ [B , \, A_i ] = - [A_i , \, B ] $.
Since $ A_i \simeq {\rm e}_i $, the standard basis vector of $ \mathbb{R}^n $, we have:
\begin{equation}
[ B , \, A_i ] \simeq b^\nu ( M_\nu ) _{ki} 
= b^\nu \begin{bmatrix} c_{\nu i}^1 & \ldots & c_{\nu i}^n \end{bmatrix}^T = -  \begin{bmatrix} c_{i\nu}^1 & \ldots & c_{i\nu}^n \end{bmatrix}^T b^\nu = - (M_i) _{k\nu} b^\nu \simeq -  [A_i , \, B ] 
\label{eq:lie-brack}
\end{equation}

\subsection{Exponential of the ${\rm ad}_{A} $ matrix}

Since the adjoint representation corresponds to a derivation of the Lie algebra, given $ A\in \mathfrak{su}(N) $
\[
 {\rm ad}_{A } = \left. \frac{d}{ d\gamma} \left( {\rm Ad}_{e^{\gamma A }} \right) \right|_{\gamma =0 } \qquad \quad \gamma \in {\mathbb{R}} 
\]
and $ {\rm ad}_{A }= M $ is the infinitesimal generator of the corresponding one-parameter group of automorphisms 
\begin{equation}
{\rm Ad}_{e^{A  \gamma }} = e^{\gamma  {\rm ad}_{A }} = e^{\gamma  M  }=  \sum_{k=0}^\infty \frac{\gamma ^k}{k!} {\rm ad}_{A }^k = \sum_{k=0}^\infty \frac{\gamma ^k}{k!}M ^k 
\label{eq:power-exp-M1}
\end{equation}
To obtain an explicit closed form expression for \eqref{eq:power-exp-M1}, we use the method of \cite{Raghunathan1} which is based on the Cayley-Hamilton theorem i.e. on expressing the exponential of a $n\times n$ matrix in terms of its first $n-1 $ powers.
In fact, if $M  $ satisfies its own characteristic equation
\begin{equation}
M  ^n = a_0 I + a_1  M  + a_2 M ^2 + \ldots + a_{n-1} M ^{n-1}
\label{eq:Cayley-Ham}
\end{equation}
then \eqref{eq:power-exp-M1} can always be written as
\begin{equation}
e^{\gamma  M } = \sum_{k=0}^{n-1} \beta_k M ^k 
\label{eq:exp-M}
\end{equation}
for suitable $ \beta_k = \beta_k ( a_0 , \, \ldots, a_{n-1} , \, \gamma  ) $.
In \cite{Raghunathan1}, the values $ \beta_k $ are obtained via a Laplace transform method that is briefly recapitulated here.
From \eqref{eq:power-exp-M1} and \eqref{eq:Cayley-Ham}, one can write $\beta_{n-1} $ as the infinite series
\begin{equation}
\beta_{n-1} = \sum_{j=0}^\infty \frac{\gamma^j}{j!} \alpha_j 
\label{eq:bm1}
\end{equation}
with $ \alpha_j $, $ j=n-1,\,n  , \, n+1 , \ldots $ satisfying the linear recurrence relation with constant coefficients
\begin{eqnarray}
\alpha_{j+1} & = & \sum_{k=0}^{n-1} a_{n-k-1} \alpha_{j-k} = a_0 \alpha_{j-n+1} + a_1 \alpha_{j-n+2} + \ldots + a_{n-1} \alpha_j \label{eq:diff-eq1}\\
\alpha_j & = & 
\begin{cases}
  0 \quad \text{ if $ j< n-1 $ } \\
  1 \quad \text{ if $ j=n-1 $} 
\end{cases}\label{eq:diff-eq2}
\end{eqnarray}
If we Laplace transform termwise $ \beta_{n-1} $, $ {\cal L}\left\{ \frac{\gamma^j}{j!} \alpha_j \right\} =  \frac{\alpha_j }{s^{j+1} }$, then the infinite sum  
\eqref{eq:bm1} becomes a $ {\cal Z}$-transform in the $ s$ variable, $ {\cal Z} (s) = \sum_{j=0}^\infty \alpha_j s^{-j} $, i.e. 
\[
\beta_{n-1} = {\cal L}^{-1} {\cal L} \left\{ \sum_{j=0}^\infty \frac{\gamma^j}{j!} \alpha_j \right\} =  {\cal L}^{-1} \left\{ \frac{1}{s}  \sum_{j=0}^\infty \alpha_j s^{-j} \right\} = {\cal L}^{-1} \left\{  \frac{1}{s} {\cal Z} (s) \right\}
\]
Multiplying both sides of \eqref{eq:diff-eq1} by $ s^{-(j+1)} $, summing for $j $ from $ n-1 $ to $ \infty$ and completing the series to zero by means of the initial conditions \eqref{eq:diff-eq2}, we obtain 
\[
\sum_{j=0}^\infty \alpha_j s^{-j} - s^{-n+1} = \left( a_0 s^{-n} + a_i s^{-n+1} + \ldots + a_{n-2} s^{-2} + a_{n-1} s^{-1} \right) \sum_{j=0}^\infty \alpha_j s^{-j} 
\]
from which, if $ s_i $, $ i=1, \ldots r \leq n$, are the eigenvalues of $ \det( sI - M  ) $ of multiplicity $ m_i $, the difference equation \eqref{eq:diff-eq1}-\eqref{eq:diff-eq2} can be written in terms of this transform as
\[
\frac{1}{s} {\cal Z}(s) = \left( s^n - a_{n-1} s^{n-1} - \ldots - a_1 s - a_0 \right)^{-1} 
= \prod_{i=1}^r  \left( s - s_i \right) ^{-m_i }
\]
or, expanding in partial fractions 
\[
\frac{1}{s} {\cal Z}(s) = \sum_{i=1}^r \sum_{j=1}^{m_i} \frac{C_{ij} }{( s - s_i )^j }
\]
where 
\[
 C_{ij} = \left. \frac{1}{( m_i-j)! } \left( \frac{d}{ d s} \right)^{m_i -j } \left({\cal Z}(s) (s-s_i )^{m_i}\right) \right|_{s=s_j}
\]
Thus antitransforming 
\begin{equation}
\beta_{n-1} = \sum_{i=1}^r \tilde C_i (\gamma ) e^{\gamma  s_i }
\label{eq:beta-n-1}
\end{equation}
with $ \tilde C_i (\gamma  ) = \sum_{j=1}^{m_i } C_{ij} \frac{\gamma  ^{j-1} } { (j-1)!} $.
The $ \beta_k $, $ k=0, \, 1 , \ldots , n-2 $, are then obtained by differentiation of $ \beta_{n-1} $ with respect to $\gamma $:
\begin{equation}
\beta_k = \left(\frac{d^{n-k-1} }{d \gamma^{n-k-1}} - a_{n-1} \frac{d^{n-k-2} }{d \gamma^{n-k-2} } - \ldots - a_{k+2} \frac{d }{d \gamma} - a_{k+1} \right) \beta_{n-1} 
\label{eq:beta-k}
\end{equation}

\subsection{Exponentials of the basis elements $ {\rm ad}_{A_i} $}
For the basis elements $ A_i \in \mathfrak{su}(N) $, the powers of the matrices $M_i = {\rm ad}_{A_i } $ are computed in terms of the structure constants in \eqref{eq:Mi-to-the-l}.
Thus, using the notation $ \beta_0^{[i]}, \beta_1^{[i]} , \ldots \beta_{n-1}^{[i]}$ for the coefficients obtained in \eqref{eq:beta-n-1}-\eqref{eq:beta-k} relative to the matrix $M_i $, we have
\begin{eqnarray}
e^{{\gamma^i \rm ad}_{A_i} } & = & \left( e^{\gamma^i M_i }\right)_{kj} = 
\beta_0 ^{[i]}\delta^k_j + \beta_1^{[i]} c_{i \, j} ^k + \beta_2^{[i]} c_{i \, \mu_1}^k c_{i \, j }^{\mu_1} + \ldots +\beta_{n-1}^{[i]} c_{i \, \mu_1}^k c_{i \, \mu_2}^{\mu_1} \ldots c_{i \, j}^{\mu_{n-2}} 
\label{eq:expMi} \\
 & = & \sum_{r=0}^{n-1} \beta_r^{[i]} c_{i \, \mu_1}^k c_{i \, \mu_2} ^{\mu_1} \ldots c_{i \, j}^{\mu_{r-1}} \nonumber
\end{eqnarray}
where it is intended that $ c_{i \, \mu_1}^k c_{i \, \mu_2} ^{\mu_1} \ldots c_{i \, j}^{\mu_{r-1}} = \delta_j^k $ for $ r=0 $ and $ c_{i \, \mu_1}^k c_{i \, \mu_2} ^{\mu_1} \ldots c_{i \, j}^{\mu_{r-1}} = c_{ij}^k $ for $r=1$ (the lower index in $ \beta_k^{[i]} $ gives the number of times the structure constants $ c_{i \, \ast}^\ast $ appear in the corresponding term).
Also the product of matrices $ e^{{\gamma^i \rm ad}_{A_i} }  $ and $ e^{{\gamma^l \rm ad}_{A_l} } $ can be expressed in terms of the $ \beta_0^{[i]}, \ldots, \beta_{n-1}^{[i]}, \beta_0^{[l]} , \ldots ,  \beta_{n-1}^{[l]} $ and of the structure constants as follows:
\begin{equation}
e^{{\gamma^i \rm ad}_{A_i} }  e^{{\gamma^l \rm ad}_{A_l} } =
\left( e^{\gamma^i M_i } e^{\gamma^l M_l }\right)_{kj} =
\sum_{r, \, s = 0 }^{n-1} \beta_r^{[i]} \beta_s^{[l]} c_{i \, \mu_1}^k c_{i \, \mu_2} ^{\mu_1} \ldots c_{i \, \nu}^{\mu_{r-1}} c_{l \, \mu_1}^\nu c_{l \, \mu_2} ^{\mu_1} \ldots c_{l \, j}^{\mu_{s-1}} 
\label{eq:expMiMl}
\end{equation}
with a similar convention as above for $ r, s = 0, 1 $.
Notice how also the ``concatenation'' of $ c_{i \, \ast } ^\ast $ and $ c_{l \, \ast }^\ast $ respects the summation convention (indeed \eqref{eq:expMiMl} represents an ordinary product of square matrices).

\section{Explicit expression of the Wei-Norman formula}
With the notation introduced in Section \ref{sec:adj}, the matrix $\Xi $ of \eqref{eq:matrix-xi} is given by 
\[
\Xi  = \begin{bmatrix}
{\rm e}_1 & e^{\gamma^1 {\rm ad}_{A_1} } {\rm e}_2 & \ldots & \prod_{i=1}^{n-1} e^{\gamma^i {\rm ad}_{A_i} } {\rm e}_n \end{bmatrix} 
\]
With the closed form expression \eqref{eq:expMi} for $ e^{\gamma^i {\rm ad}_{A_i}} $, it is possible to compute $ \Xi $ explicitly without any infinite summation.
Looking at \eqref{eq:wei-tmp1} and \eqref{eq:expMiMl} it is clear what we have to do:
from the matrix products of the $ e^{\gamma^i {\rm ad}_{A_i} }$ compute all the column vectors corresponding to $ \prod_{i=1}^{j-1} e^{\gamma^i {\rm ad}_{A_i} } A_j $, $j =2, \ldots , n $, in the same fashion as \eqref{eq:lie-brack} and then regroup them along the $n$ basis directions $A_k $.
This can be done explicitly.
\begin{equation}
\begin{split}
e^{\gamma^1 {\rm ad}_{A_1} } A_2 & \simeq \sum_{r=0}^{n-1} \beta_r^{[1]} c_{1 \, \mu_1}^k  \ldots c_{1 \, j}^{\mu_{r-1}} \\
e^{\gamma^1 {\rm ad}_{A_1} } e^{\gamma^2 {\rm ad}_{A_2} } A_3 & \simeq  
\sum_{r_1, r_2 =0}^{n-1} \beta_{r_1}^{[1]}\beta_{r_2}^{[2]} c_{1 \, \mu_1}^k  \ldots c_{1 \, \nu}^{\mu_{r_1-1}} c_{2 \, \mu_1}^\nu  \ldots c_{2 \, 3}^{\mu_{r_2-1}} \\
 & \vdots \\
\prod_{i=1}^{n-1} e^{\gamma^i {\rm ad}_{A_i} } A_n & \simeq \sum_{r_1, \ldots , r_{n-1} = 0 }^{n-1}  \beta_{r_1}^{[1]} \ldots \beta_{r_{n-1}}^{[n-1]}  c_{1 \, \mu_1}^k  \ldots c_{1 \, \nu}^{\mu_{r_1-1}} c_{2 \, \mu_1}^\nu  \ldots \ldots c_{n-2 \, \nu}^{\mu_{r_{n-2}-1}} c_{n-1 \, \mu_1}^\nu \ldots c_{n-1 \, n}^{\mu_{r_{n-1}-1}} 
\end{split}
\label{eq:wei-adj}
\end{equation}
The only free index in \eqref{eq:wei-adj} is $k$ i.e. each $ \prod_{i=1}^{j-1} e^{\gamma^i {\rm ad}_{A_i} } A_j $ is an $n$-dimensional vector as expected.
The expression of the $n\times n$ matrix $ \Xi $ one obtains by stacking together the columns vectors of \eqref{eq:wei-adj} is
\begin{equation}
\Xi = \begin{bmatrix}
1 & \sum_{r=0}^{n-1} \beta_r^{[1]} c_{1 \, \mu_1}^1  \ldots c_{1 \, j}^{\mu_{r-1}}
& \ldots & \sum_{r_1, \ldots , r_{n-1} = 0 }^{n-1}  \beta_{r_1}^{[1]} \ldots \beta_{r_{n-1}}^{[n-1]}  c_{1 \, \mu_1}^1  \ldots c_{n-1 \, n}^{\mu_{r_{n-1}-1}} \\
0 &  \sum_{r=0}^{n-1} \beta_r^{[1]} c_{1 \, \mu_1}^2  \ldots c_{1 \, j}^{\mu_{r-1}}
& \ldots & \sum_{r_1, \ldots , r_{n-1} = 0 }^{n-1}  \beta_{r_1}^{[1]} \ldots \beta_{r_{n-1}}^{[n-1]}  c_{1 \, \mu_1}^2  \ldots c_{n-1 \, n}^{\mu_{r_{n-1}-1}} \\
\vdots & \vdots & \vdots & \vdots \\
0 &  \sum_{r=0}^{n-1} \beta_r^{[1]} c_{1 \, \mu_1}^n  \ldots c_{1 \, j}^{\mu_{r-1}}
& \ldots & \sum_{r_1, \ldots , r_{n-1} = 0 }^{n-1}  \beta_{r_1}^{[1]} \ldots \beta_{r_{n-1}}^{[n-1]}  c_{1 \, \mu_1}^n  \ldots c_{n-1 \, n}^{\mu_{r_{n-1}-1}}
\end{bmatrix}
\label{eq:wei-n-matrix}
\end{equation}

\section{Example: $\mathfrak{su}(2)$}
\label{sec:su2}
A skew-symmetric basis for $ \mathfrak{su}(2) $ is obtained from the Pauli matrices
\[
\sigma_1 =  \begin{bmatrix} 0 & 1 \\ 1 & 0 \end{bmatrix} \quad
\sigma_2 =  \begin{bmatrix} 0 & -i \\ i & 0 \end{bmatrix} \quad
\sigma_3 =  \begin{bmatrix} 1 & 0 \\ 0 & -1 \end{bmatrix}
\]
for example by taking $ A_j =\frac{i}{2} \sigma_j $, $j=1,\,2,\,3$, i.e.
\begin{equation}
A_1 = \frac{1}{2} \begin{bmatrix} 0 & i \\ i & 0 \end{bmatrix} \quad
A_2 = \frac{1}{2} \begin{bmatrix} 0 & 1 \\ -1 & 0 \end{bmatrix} \quad
A_3 = \frac{1}{2} \begin{bmatrix} i & 0 \\ 0 & -i \end{bmatrix}
\label{eq:basis-su2}
\end{equation}
and it corresponds to all real structure constants $ c_{12}^3 = c_{23}^1 = c_{31}^2 = 1$.
The corresponding adjoint matrices are
\[
{\rm ad}_{A_1} =  \begin{bmatrix}
0 & 0 & 0 \\
0 & 0 & -1 \\
0 & 1 & 0 
\end{bmatrix} \quad
{\rm ad}_{A_2} =  \begin{bmatrix}
0 & 0 & 1 \\
0 & 0 & 0 \\
-1 & 0 & 0 
\end{bmatrix} \quad
{\rm ad}_{A_3} =  \begin{bmatrix}
0 & -1 & 0 \\
1 & 0 & 0 \\
0 & 0 & 0 
\end{bmatrix} \quad
\]
whose exponentials are already known from the literature to be:
\[
e^{\gamma^1 {\rm ad}_{A_1} } =
\begin{bmatrix}
1 & 0 & 0 \cr 
0 & \cos (\gamma^{1}) & -\sin (\gamma^{1}) \cr 
0 & \sin (\gamma^{1}) & \cos (\gamma^{1}) \cr  
\end{bmatrix} \quad 
e^{\gamma^2 {\rm ad}_{A_2} } =\begin{bmatrix}
\cos (\gamma^{2}) & 0 & \sin (\gamma^{2}) \cr 
0 & 1 & 0 \cr 
-\sin (\gamma^{2}) & 0 & \cos (\gamma^{2}) \cr  
\end{bmatrix} \quad 
e^{\gamma^3 {\rm ad}_{A_3} } =\begin{bmatrix}
\cos (\gamma^{3}) & -\sin (\gamma^{3}) & 0 \cr
 \sin (\gamma^{3}) & \cos ( \gamma^{3}) & 0 \cr 
0 & 0 & 1 \cr 
\end{bmatrix}
\]
The Magnus expansion \eqref{eq:Magnus-expan1} is given by $ U(t) = T  \int _0^t  e^{u^1 A_1 + u^2 A_2 + u^3 A_3} d\tau  $ (with $ u^i = u^i (t) $).

\subsection{Wei-Norman formula for canonical coordinates of the second kind}
\label{sec:su2-a}
In the product of exponentials, choosing the order given by the cardinality of the index gives the canonical coordinates of the second kind on $ SU(2)$:
$ U(t) = e^{\gamma^1 A_1}e^{ \gamma^2 A_2}e^{ \gamma^3 A_3} $ (again with $ \gamma^i = \gamma^i (t) $).
For this choice, the Wei-Norman formula \eqref{eq:wei-n-matrix} reads as:
\begin{equation}
\Xi =\begin{bmatrix}
1 & 0 &  \sin (\gamma^{2}) \cr 
0 & \cos (\gamma^{1}) & - \cos (\gamma^{2})\,
   \sin (\gamma^{1}) \cr 
0 & \sin (\gamma^{1}) & \cos (\gamma^{1})\,
   \cos (\gamma^{2}) \cr 
\end{bmatrix}
\label{eq:wei-n-su2}
\end{equation}
whose inverse can be computed explicitly:
\begin{equation}
\Xi^{-1} =\begin{bmatrix}
 1 & \sin (\gamma^{1})\,\tan (\gamma^{2}) & - \cos (\gamma^{1})\,
     \tan (\gamma^{2})  \cr 
0 & \cos (\gamma^{1}) & \sin (\gamma^{1}) \cr 
0 & - \sec (\gamma^{2})\,\sin (\gamma^{1})  & \cos (\gamma^{1})\,\sec (\gamma^{2}) \cr 
\end{bmatrix}
\label{eq:wei-n-su2-inv}
\end{equation}
From \eqref{eq:wei-n-su2}, the determinant of $ \Xi $ is simply
\[
\det \Xi = \cos \gamma^2
\]
and thus the singularities of the representation are $ \gamma^2 = \pi/2 + k \pi $, $ k \in \mathbb{Z} $.
While \eqref{eq:matrix-xi} (and \eqref{eq:wei-n-su2}) is valid everywhere, in the singular points the formula \eqref{eq:matrix-xiinv} cannot be applied (i.e. \eqref{eq:wei-n-su2-inv} is not defined).

\subsection{Wei-Norman formula for the $ZYZ $ Euler angles}
When expressed in the basis \eqref{eq:basis-su2}, the ZYZ Euler angles correspond to the product of exponentials $ U(t) = e^{\gamma^1 A_3}e^{ \gamma^2 A_2}e^{ \gamma^3 A_3} $ (compare with the expression of Section \ref{sec:su2-a}).
The Wei-Norman formula corresponds in this case to 
\[
\Xi = \begin{bmatrix} {\rm e}_3 \; & \; e^{\gamma^1 {\rm ad}_{A_3} } {\rm e}_2 \;  & \;  
 e^{\gamma^1 {\rm ad}_{A_3} } e^{\gamma^2 {\rm ad}_{A_2} } {\rm e}_3 
\end{bmatrix} =
 \begin{bmatrix} 0 & - \sin ( \gamma^1 ) & \cos ( \gamma^1 ) \sin ( \gamma^2 ) \\
0 & \cos ( \gamma^1) & \sin ( \gamma^1 ) \sin ( \gamma^2 ) \\
1 & 0 & \cos ( \gamma^2 ) 
\end{bmatrix}
\]
and its inverse
\[ 
\Xi^{-1} = 
 \begin{bmatrix} - \cos ( \gamma^1 )  \cot ( \gamma^2 ) & - \sin ( \gamma^1 ) \cot ( \gamma^2 ) & 1 \\
-  \sin ( \gamma^1) & \cos ( \gamma^1 ) & 0 \\
\cos ( \gamma^1 ) \csc ( \gamma^2 ) & \sin ( \gamma^1 ) \csc (\gamma^2 ) & 0 
\end{bmatrix}
\]
Since 
\[
 \det(\Xi ) = \sin \gamma^2
\]
the singularity has now moved to $  \gamma^2 = k \pi $, $ k \in \mathbb{Z} $, as is well-known for such a parameterization.
Thus $ \Xi^{-1}$ can be used everywhere except in the identity $ U(0) = I $.
It is worth emphasizing that it is a fundamental topological fact that singularities cannot be avoided in a minimal parameterization of a semisimple Lie group.
One possible way to get around the problem is obviously to use ``redundant'' parameterizations like quaternions.

\section{Example: $\mathfrak{su}(3)$}
The following skew-Hermitian basis obtained for example by considering $ \mathfrak{su}(3) $ as the compact real form of the classical $ A_2 $ algebra with its Cartan basis is more suited for the analysis of three level systems than the Gell-Mann basis normally used in particle physics.
\begin{equation}
\begin{split}
i H_1 = \begin{bmatrix} i & 0 & 0 \\ 0 & -i & 0 \\ 0 & 0 & 0 \end{bmatrix}
\quad 
i H_2 = \begin{bmatrix} 0 & 0 & 0 \\ 0 & i & 0 \\ 0 & 0 & -i\end{bmatrix} 
\quad 
X_{12} = \begin{bmatrix} 0 & 1 & 0 \\ -1 & 0 & 0 \\ 0 & 0 & 0 \end{bmatrix}
\quad 
Y_{12} = \begin{bmatrix} 0 & i & 0 \\ i & 0 & 0 \\ 0 & 0 & 0 \end{bmatrix}
\\
X_{13} = \begin{bmatrix} 0 & 0 & 1 \\ 0 & 0 & 0 \\ -1 & 0 & 0 \end{bmatrix}
\quad 
Y_{13} = \begin{bmatrix} 0 & 0 & i \\ 0 & 0 & 0 \\ i & 0 & 0 \end{bmatrix}
\quad 
X_{23} = \begin{bmatrix} 0 & 0 & 0 \\ 0 & 0 & 1 \\ 0 & -1 & 0 \end{bmatrix}
\quad 
Y_{23} = \begin{bmatrix} 0 & 0 & 0 \\ 0 & 0 & i \\ 0 & i & 0 \end{bmatrix}
\end{split}
\label{eq:basis-su3}
\end{equation}
The nonnull structure constants $ c^k_{ij} = - c^k_{ji}$ are:
\begin{equation}
\begin{split}
c_{13}^4=c_{41}^3 = c_{34}^1= 2, \quad
c_{15}^6=c_{61}^5 = c_{56}^1= 2, \quad
c_{27}^8=c_{82}^7 = c_{78}^2= 2 \\
c_{45}^8=c_{84}^5 = c_{58}^4= 1, \quad
c_{46}^7=c_{74}^6 = c_{67}^4= -1, \quad
c_{35}^7=c_{73}^5 = c_{57}^3= -1, \quad
c_{36}^8=c_{83}^6 = c_{68}^3= -1 \\
c_{17}^8=c_{81}^7 = -1, \quad c_{78}^1= 0, \quad
c_{23}^4=c_{42}^3 = -1, \quad c_{34}^2= 0, \quad
c_{25}^6=c_{62}^5 = 1, \quad c_{56}^2= 2 
\label{eq:struct-const-su3}
\end{split}
\end{equation}
Note that because of the last row of \eqref{eq:struct-const-su3} the presentation of $ {\rm ad}_A $ is not skew-symmetric, like it would have been had we chosen the Gell-Mann basis.

For $\mathfrak{su}(3) $, the basis of the adjoint representation is
\[
M_1 =
\begin{bmatrix}
 0 & 0 & 0 & 0 & 0 & 0 & 0 & 0 \cr 0 & 0 & 0 & 0 & 0 & 0 & 0 & 0 \cr \
0 & 0 & 0 & c_{14}^{3} & 0 & 0 & 0 & 0 \cr 0 & 0 & c_{13}^{4} & 0 \
& 0 & 0 & 0 & 0 \cr 0 & 0 & 0 & 0 & 0 & c_{16}^{5} & 0 & 0 \cr 0 & 0 & \
0 & 0 & c_{15}^{6} & 0 & 0 & 0 \cr 0 & 0 & 0 & 0 & 0 & 0 & 0 & \
c_{18}^{7} \cr 0 & 0 & 0 & 0 & 0 & 0 & c_{17}^{8} & 0 \cr  
\end{bmatrix}, \quad 
M_2 = \begin{bmatrix} 
0 & 0 & 0 & 0 & 0 & 0 & 0 & 0 \cr 0 & 0 & 0 & 0 & 0 & 0 & 0 & 0 \cr \
0 & 0 & 0 & c_{24}^{3} & 0 & 0 & 0 & 0 \cr 0 & 0 & c_{23}^{4} & 0 \
& 0 & 0 & 0 & 0 \cr 0 & 0 & 0 & 0 & 0 & c_{26}^{5} & 0 & 0 \cr 0 & 0 & \
0 & 0 & c_{25}^{6} & 0 & 0 & 0 \cr 0 & 0 & 0 & 0 & 0 & 0 & 0 & \
c_{28}^{7} \cr 0 & 0 & 0 & 0 & 0 & 0 & c_{27}^{8} & 0 \cr
\end{bmatrix}
\]
\[
M_3 =\begin{bmatrix} 
0 & 0 & 0 & c_{34}^{1} & 0 & 0 & 0 & 0 \cr 0 & 0 & 0 & 0 & 0 & 0 \
& 0 & 0 \cr 0 & 0 & 0 & 0 & 0 & 0 & 0 & 0 \cr c_{31}^{4} & \
c_{32}^{4} & 0 & 0 & 0 & 0 & 0 & 0 \cr 0 & 0 & 0 & 0 & 0 & 0 & \
c_{37}^{5} & 0 \cr 0 & 0 & 0 & 0 & 0 & 0 & 0 & c_{38}^{6} \cr 0 & \
0 & 0 & 0 & c_{35}^{7} & 0 & 0 & 0 \cr 0 & 0 & 0 & 0 & 0 & \
c_{36}^{8} & 0 & 0 \cr  
\end{bmatrix}, \quad 
M_4 =  
\begin{bmatrix}
0 & 0 & c_{43}^{1} & 0 & 0 & 0 & 0 & 0 \cr 0 & 0 & 0 & 0 & 0 & 0 \
& 0 & 0 \cr c_{41}^{3} & c_{42}^{3} & 0 & 0 & 0 & 0 & 0 & 0 \cr 0 \
& 0 & 0 & 0 & 0 & 0 & 0 & 0 \cr 0 & 0 & 0 & 0 & 0 & 0 & 0 & c_{48}^{5} \
\cr 0 & 0 & 0 & 0 & 0 & 0 & c_{47}^{6} & 0 \cr 0 & 0 & 0 & 0 & 0 & \
c_{46}^{7} & 0 & 0 \cr 0 & 0 & 0 & 0 & c_{45}^{8} & 0 & 0 & 0 \cr \
 \end{bmatrix} 
\]
\[
M_5=\begin{bmatrix}
0 & 0 & 0 & 0 & 0 & c_{56}^{1} & 0 & 0 \cr 0 & 0 & 0 & 0 & 0 & \
c_{56}^{2} & 0 & 0 \cr 0 & 0 & 0 & 0 & 0 & 0 & c_{57}^{3} & 0 \cr \
0 & 0 & 0 & 0 & 0 & 0 & 0 & c_{58}^{4} \cr 0 & 0 & 0 & 0 & 0 & 0 & 0 & \
0 \cr c_{51}^{6} & c_{52}^{6} & 0 & 0 & 0 & 0 & 0 & 0 \cr 0 & 0 & \
c_{53}^{7} & 0 & 0 & 0 & 0 & 0 \cr 0 & 0 & 0 & c_{54}^{8} & 0 & 0 \
& 0 & 0 \cr  
\end{bmatrix}, \quad 
M_6 =
\begin{bmatrix}
0 & 0 & 0 & 0 & c_{65}^{1} & 0 & 0 & 0 \cr 0 & 0 & 0 & 0 & \
c_{65}^{2} & 0 & 0 & 0 \cr 0 & 0 & 0 & 0 & 0 & 0 & 0 & c_{68}^{3} \
\cr 0 & 0 & 0 & 0 & 0 & 0 & c_{67}^{4} & 0 \cr c_{61}^{5} & \
c_{62}^{5} & 0 & 0 & 0 & 0 & 0 & 0 \cr 0 & 0 & 0 & 0 & 0 & 0 & 0 & 0 \
\cr 0 & 0 & 0 & c_{64}^{7} & 0 & 0 & 0 & 0 \cr 0 & 0 & c_{63}^{8} \
& 0 & 0 & 0 & 0 & 0 \cr  
\end{bmatrix}
\]
\[
M_7=
\begin{bmatrix} 0 & 0 & 0 & 0 & 0 & 0 & 0 & 0 \cr 0 & 0 & 0 & 0 & 0 & 0 & 0 & \
c_{78}^{2} \cr 0 & 0 & 0 & 0 & c_{75}^{3} & 0 & 0 & 0 \cr 0 & 0 & \
0 & 0 & 0 & c_{76}^{4} & 0 & 0 \cr 0 & 0 & c_{73}^{5} & 0 & 0 & 0 \
& 0 & 0 \cr 0 & 0 & 0 & c_{74}^{6} & 0 & 0 & 0 & 0 \
\cr 0 & 0 & 0 & 0 & 0 & 0 & 0 & 0 \cr c_{71}^{8} & c_{72}^{8} & 0 \
& 0 & 0 & 0 & 0 & 0 \cr 
\end{bmatrix}, \quad 
M_8 =
\begin{bmatrix}
0 & 0 & 0 & 0 & 0 & 0 & 0 & 0 \cr 0 & 0 & 0 & 0 & 0 & 0 & \
c_{87}^{2} & 0 \cr 0 & 0 & 0 & 0 & 0 & c_{86}^{3} & 0 & 0 \cr 0 & \
0 & 0 & 0 & c_{85}^{4} & 0 & 0 & 0 \cr 0 & 0 & 0 & c_{84}^{5} & 0 \
& 0 & 0 & 0 \cr 0 & 0 & c_{83}^{6} & 0 & 0 & 0 & 0 & \
0 \cr c_{81}^{7} & c_{82}^{7} & 0 & 0 & 0 & 0 & 0 & 0 \cr 0 & 0 & \
0 & 0 & 0 & 0 & 0 & 0 \cr 
\end{bmatrix}
\]

The characteristic polynomials and corresponding eigenvalues of the basis elements $ M_i $, $ i=1, \ldots, m$ are: 
\begin{equation}
\det (sI - M_i ) =
\begin{cases}
s^8 +9 s^6 + 24s^4 +   16 s^2  \qquad \text{ for $i=1 $} \\
s^8 +6 s^6 + 9s^4 +   4 s^2 \qquad \text{ for $i=2,\, 3 , \, 4, \, 7, \, 8 $} \\
s^8 +8 s^6 + 13s^4 +   6 s^2 \qquad \text{ for $i=5, \, 6 $} 
\end{cases}
\label{eq:chp-su3}
\end{equation}
\[
{\rm eig}(M_i) =
\begin{cases}
\left\{ 0, \, 0, \, \pm i , \, \pm 2 i , \,  \pm 2 i\right\} \qquad \text{ for $i=1 $} \\
 \left\{ 0, \, 0, \, \pm i , \, \pm  i , \,  \pm 2 i\right\}\qquad \text{ for $i=2,\, 3 , \, 4, \, 7, \, 8 $} \\
 \left\{ 0, \, 0, \, \pm i , \, \pm  i , \,  \pm \sqrt{6} i\right\} \qquad \text{ for $i=5, \, 6 $} 
\end{cases}
\]
Inserting the coefficients of the characteristic polynomials \eqref{eq:chp-su3} into \eqref{eq:Cayley-Ham}, one gets from \eqref{eq:beta-n-1} and \eqref{eq:beta-k} the following values for the $ \beta^{[i]}_k $, $k=0, \ldots , 7$ and $ i=1, \ldots, 8$: 
\begin{eqnarray*}
\beta_0^{[i]} & = & 1 \qquad \text{ for $i=1, \ldots, 8 $} \\
\beta_1^{[i]} & = & \gamma^i \qquad \text{ for $i=1, \ldots, 8 $} \\
\beta_2^{[i]} & = & 
\begin{cases}
\frac{1} {36}
\left({54 - 64\,\cos (\gamma^i) + 10\,\cos (2\,\gamma^i) + 3\,\gamma^i\,\sin (2\,\gamma^i)} \right) &
 \qquad \text{ for $i=1 $} \\
\frac{1} {36} 
\left({81 - 80\,\cos (\gamma^i) - {\cos^2 (\gamma^i)} - 24\,\gamma^i\,\sin (\gamma^i) + 
    \sin^2 (\gamma^i)} \right) &
\qquad \text{ for $i=2,\, 3 , \, 4, \, 7, \, 8 $} \\
\frac{1} {150}
\left({325 - 324\,\cos (\gamma^i) - \cos ({\sqrt{6}}\,\gamma^i) - 90\,\gamma^i\,\sin (\gamma^i)} \right) &
\qquad \text{ for $i=5, \, 6 $} 
\end{cases}
\\
\beta_3^{[i]} & = &
\begin{cases}
\frac{1} {144}
\left({216\,\gamma^i - 6\,\gamma^i\,\cos (2\,\gamma^i) - 256\,\sin (\gamma^i) + 
    23\,\sin (2\,\gamma^i)} \right) &
 \qquad \text{ for $i=1 $} \\
\frac{1}  {36} 
\left({81\,\gamma^i + 24\,\gamma^i\,\cos (\gamma^i) - 104\,\sin (\gamma^i) - 
    \cos (\gamma^i)\,\sin (\gamma^i)}\right) & 
\qquad \text{ for $i=2,\, 3 , \, 4, \, 7, \, 8 $} \\
\frac{13\,\gamma^i}{6} + \frac{3\,\gamma^i\,\cos (\gamma^i)}{5} - \frac{69\,\sin (\gamma^i)}{25} - 
  \frac{\sin ({\sqrt{6}}\,\gamma^i)}{150\,{\sqrt{6}}} & 
\qquad \text{ for $i=5, \, 6 $} 
\end{cases}
 \\
\beta_4^{[i]} & = & \begin{cases}
\frac{1}  {144}
\left({81 + 47\,\cos (2\,\gamma^i) + \cos (\gamma^i)\,\left( -128 + 30\,\gamma^i\,\sin (\gamma^i) \right) } \right) &
 \qquad \text{ for $i=1 $} \\
\frac{1}{18}
\left({27 - 26\,\cos (\gamma^i) - {\cos^2 (\gamma^i)} - 15\,\gamma^i\,\sin (\gamma^i) + 
    {\sin^2 (\gamma^i)}} \right) &
\qquad \text{ for $i=2,\, 3 , \, 4, \, 7, \, 8 $} \\
\frac{1} {150}
\left(
{200 - 198\,\cos (\gamma^i) - 2\,\cos ({\sqrt{6}}\,\gamma^i) - 105\,\gamma^i\,\sin (\gamma^i)} \right) & 
\qquad \text{ for $i=5, \, 6 $} 
\end{cases}
\\
\beta_5^{[i]} & = & 
\begin{cases}
\frac{1}{576}
\left( {324\,\gamma^i - 30\,\gamma^i\,\cos (2\,\gamma^i) - 512\,\sin (\gamma^i) + 
    109\,\sin (2\,\gamma^i)} \right) &
 \qquad \text{ for $i=1 $} \\
\frac{1}{18}
\left({27\,\gamma^i + 15\,\gamma^i\,\cos (\gamma^i) - 41\,\sin (\gamma^i) - 
    \cos (\gamma^i)\,\sin (\gamma^i)} \right)  &
\qquad \text{ for $i=2,\, 3 , \, 4, \, 7, \, 8 $} \\
\frac{1}{450}
\left({600\,\gamma^i + 315\,\gamma^i\,\cos (\gamma^i) - 909\,\sin (\gamma^i) - 
    {\sqrt{6}}\,\sin ({\sqrt{6}}\,\gamma^i)} \right) &
\qquad \text{ for $i=5, \, 6 $} 
\end{cases}
\\
\beta_6^{[i]} & = & 
\begin{cases}
\frac{1}{144} 
\left({9 + 7\,\cos (2\,\gamma^i) + \cos (\gamma^i)\,\left( -16 + 6\,\gamma^i\,\sin (\gamma^i) \right) } \right) &
 \qquad \text{ for $i=1 $} \\
\frac{1}{36}\left( 9 - 8\,\cos (\gamma^i) - {\cos^2 (\gamma^i)} - 6\,\gamma^i\,\sin (\gamma^i) +    {\sin^2 (\gamma^i)} \right) &
\qquad \text{ for $i=2,\, 3 , \, 4, \, 7, \, 8 $} \\
\frac{1}{150}
\left(25 - 24\,\cos (\gamma^i) - \cos ({\sqrt{6}}\,\gamma^i) - 15\,\gamma^i\,\sin (\gamma^i)\right) &
\qquad \text{ for $i=5, \, 6 $} 
\end{cases}
\\
\beta_7^{[i]} & = & 
\begin{cases}
\frac{1}{576}
\left(36\,\gamma^i - 6\,\gamma^i\,\cos (2\,\gamma^i) - 64\,\sin (\gamma^i) + 17\,\sin (2\,\gamma^i) \right) &
 \qquad \text{ for $i=1 $} \\
\frac{1}{36}
\left(9\,\gamma^i + 6\,\gamma^i\,\cos (\gamma^i) - 14\,\sin (\gamma^i) - 
    \cos (\gamma^i)\,\sin (\gamma^i)\right) &
\qquad \text{ for $i=2,\, 3 , \, 4, \, 7, \, 8 $} \\
\frac{1}{900}
\left(150\,\gamma^i + 90\,\gamma^i\,\cos (\gamma^i) - 234\,\sin (\gamma^i) - 
    {\sqrt{6}}\,\sin ({\sqrt{6}}\,\gamma^i) \right) &
\qquad \text{ for $i=5, \, 6 $} 
\end{cases}
\end{eqnarray*}
and, from \eqref{eq:exp-M}, the exponentials are:
\[
e^{\gamma_1 {\rm ad}_{A_1}} =
\begin{bmatrix}
 1 &  0 &  0 &  0 &  0 &  0 &  0 &  0 \\
 0 &  1 &  0 &  0 &  0 &  0 &  0 &  0 \\
 0 &  0 & \cos (2 \gamma^1 ) & -2 \cos (\gamma^1) \sin(\gamma^1) &  0 &  0 &  0 &  0 \\
0 &  0 & \sin ( 2 \gamma^1 ) & \cos (2 \gamma^1 ) &  0 &  0 &  0 &  0 \\
 0 &  0 &  0 &  0 & \cos (2 \gamma^1 ) &-2 \cos (\gamma^1) \sin (\gamma^1) &  0 &  0 \\
 0 &  0 &  0 &  0 & \sin ( 2 \gamma^1 ) & \cos (2 \gamma^1 ) &  0 &  0 \\
 0 &  0 &  0 &  0 &  0 &  0 & \cos (\gamma^1 )& \sin (\gamma^1) \\
 0 &  0 &  0 &  0 &  0 &  0 & - \sin (\gamma^1) & \cos (\gamma^1) 
\end{bmatrix}	
\]
\[
 e^{\gamma^2 {\rm ad}_{A_2}} =
\begin{bmatrix}
1 & 0 & 0 & 0 & 0 & 0 & 0 & 0 \cr 
0 & 1 & 0 & 0 & 0 & 0 & 0 & 0 \cr 
0 & 0 & \cos (\gamma^2) & \sin (\gamma^2) & 0 & 0 & 0 & 0 \cr 
0 & 0 & - \sin (\gamma^2) & \cos (\gamma^2) & 0 & 0 & 0 & 0 \cr 
0 & 0 & 0 & 0 & \cos (\gamma^2) & - \sin (\gamma^2) & 0 & 0 \cr 
0 & 0 & 0 & 0 & \sin (\gamma^2) & \cos (\gamma^2) & 0 & 0 \cr 
0 & 0 & 0 & 0 & 0 & 0 & \cos (2  \gamma^2) & - \sin (2 \gamma^2) \cr 
0 & 0 & 0 & 0 & 0 & 0 & \sin (2 \gamma^2) & \cos (2 \gamma^2) \cr
\end{bmatrix}	
\]
\[
 e^{\gamma^3 {\rm ad}_{A_3}} =
\begin{bmatrix}
\cos (2 \gamma^3) & \sin^2 (\gamma^3) & 0 & 
  \sin (   2\gamma^3) & 0 & 0 & 0 & 0 \cr 
0 & 1 & 0 & 0 & 0 & 0 & 0 & 0 \cr 
0 & 0 & 1 & 0 &  0 & 0 & 0 & 0 \cr -\sin (2\gamma^3) & 
  \cos (\gamma^3) \sin (\gamma^3) & 0 & \cos (  2\gamma^3) & 0 & 0 & 0 & 0 \cr 
0 & 0 & 0 & 0 & \cos (\gamma^3) & 0 & \sin (\gamma^3 )& 0 \cr 
0 & 0 & 0 & 0 & 0 & \cos (\gamma^3) & 0 & \sin (\gamma^3) \cr 
0 & 0 & 0 & 0 & -\sin (\gamma^3) & 0 & \cos (\gamma^3) & 0 \cr 
0 & 0 & 0 & 0 & 0 & -\sin (\gamma^3) & 0 & \cos (\gamma^3) \cr 
\end{bmatrix}	
\]
\[
e^{\gamma^4 {\rm ad}_{A_4}} =
\begin{bmatrix}
\cos (2\,\gamma^4) & \sin^2 (\gamma^4) & 
 -\sin (2\gamma^4) & 0 & 0 & 0 & 0 & 0 \cr 
0 & 1 & 0 & 0 & 0 & 0 & 0 & 0 \cr 
\sin (   2\gamma^4) & - \cos (\gamma^4) \,\sin( \gamma^4)  & 
  \cos ( 2\gamma^4) & 0 & 0 & 0 & 0 & 0 \cr 
0 & 0 & 0 & 1 & 0 & 0 & 0 & 0 \cr 
0 & 0 & 0 &    0 & \cos (\gamma^4) & 0 & 0 & -\sin (\gamma^4) \cr 
0 & 0 & 0 & 0 & 0 & \cos (\gamma^4) & \sin  (\gamma^4) & 0 \cr 
0 & 0 & 0 & 0 & 0 & -\sin (\gamma^4) & \cos  (\gamma^4) & 0 \cr 
0 & 0 & 0 & 0 & \sin (\gamma^4) & 0 & 0 & \cos (\gamma^4) 
\end{bmatrix}	
\]
\[
 e^{\gamma^5 {\rm ad}_{A_5}} =
\begin{bmatrix}
\frac{1 + 2\,\cos ({\sqrt{6}}\,\gamma^5)}{3} & 
 \frac{-1 +   \cos ({\sqrt{6}}\,\gamma^5)}{3} & 0 & 0 & 0 & 
 {\sqrt{\frac{2}{3}}}\, \sin ({\sqrt{6}}\,\gamma^5) & 0 & 0 \cr 
\frac{2\, \left( -1 + \cos ({\sqrt{6}}\,\gamma^5) \right) }{3} & 
 \frac{2 +  \cos ({\sqrt{6}}\,\gamma^5)}{3} & 0 & 0 & 0 & 
 {\sqrt{\frac{2}{3}}}\,   \sin ({\sqrt{6}}\,\gamma^5) & 0 & 0 \cr 
0 & 0 & \cos (\gamma^5) & 0 & 0 & 0 & -\sin (\gamma^5) & 0 \cr 
0 & 0 & 0 & \cos (\gamma^5) & 0 & 0 & 0 & \sin (\gamma^5) \cr 
0 & 0 & 0 & 0 & 1 & 0 & 0 & 0 \cr 
-\left( {\sqrt{\frac{2}{3}}}\, \sin ({\sqrt{6}}\,\gamma^5) \right)  & 
 -\left( \frac{\sin ({\sqrt{6}}\,\gamma^5)} {{\sqrt{6}}} \right)  & 0 & 
 0 & 0 & \cos ({\sqrt{6}}\,  \gamma^5) & 0 & 0 \cr 
0 & 0 & \sin (\gamma^5) & 0 & 0 & 0 & \cos  (\gamma^5) & 0 \cr 
0 & 0 & 0 & -\sin (\gamma^5) & 0 & 0 & 0 & \cos (\gamma^5)
\end{bmatrix}	
\]
\[
 e^{\gamma^6 {\rm ad}_{A_6}} =
\begin{bmatrix}
\frac{1 + 2\,\cos ({\sqrt{6}}\,\gamma^6)}{3} & 
 \frac{-1 + \cos ({\sqrt{6}}\,\gamma^6)}{3} & 0 & 0 & 
 -\left( {\sqrt{\frac{2}{3}}}\, \sin ({\sqrt{6}}\,\gamma^6) \right)  & 
 0 & 0 & 0 \cr 
\frac{2\, \left( -1 + \cos ({\sqrt{6}}\,\gamma^6) \right) }{3} & 
 \frac{2 + \cos ({\sqrt{6}}\,\gamma^6)}{3} & 0 & 0 & 
 -\left( {\sqrt{\frac{2}{3}}}\, \sin ({\sqrt{6}}\,\gamma^6) \right)  & 
 0 & 0 & 0 \cr 
0 & 0 & \cos (\gamma^6) & 0 & 0 & 0 & 0 & -\sin (\gamma^6) \cr 0 & 0 & 0 & 
 \cos (\gamma^6) & 0 & 0 & -\sin (\gamma^6) & 0 \cr 
{\sqrt{\frac{2}{3}}}\,\sin ({\sqrt{6}}\,\gamma^6) & 
\frac{\sin ( {\sqrt{6}}\,\gamma^6)}{{\sqrt{6}}} & 0 & 0 & 
 \cos ({\sqrt{6}}\, \gamma^6) & 0 & 0 & 0 \cr 
0 & 0 & 0 & 0 & 0 & 1 & 0 & 0 \cr 
0 & 0 & 0 & \sin (\gamma^6) & 0 & 0 & \cos (\gamma^6) & 0 \cr 
0 & 0 & \sin (\gamma^6) & 0 & 0 & 0 & 0 & \cos (\gamma^6)
\end{bmatrix}	
\]
\[
 e^{\gamma^7 {\rm ad}_{A_7}} =
\begin{bmatrix}
1 & 0 & 0 & 0 & 0 & 0 & 0 & 0 \cr 
\sin^2 (\gamma^7) & \cos (  2\,\gamma^7) & 0 & 0 & 0 & 0 & 0 & 
 \sin (2\,\gamma^7) \cr 
0 & 0 & \cos (\gamma^7) & 0 & \sin  (\gamma^7) & 0 & 0 & 0 \cr 
0 & 0 & 0 & \cos (\gamma^7) & 0 & \sin (\gamma^7) & 0 & 0 \cr 
0 & 0 & -\sin (\gamma^7) & 0 & \cos (\gamma^7) & 0 & 0 & 0 \cr 
0 & 0 & 0 & -\sin (\gamma^7) & 0 & \cos (\gamma^7) & 0 & 0 \cr 
0 & 0 & 0 & 0 & 0 & 0 & 1 & 0 \cr 
\cos (\gamma^7)\,\sin (\gamma^7) & -\sin (2\,\gamma^7) & 0 & 0 & 
 0 & 0 & 0 & \cos (2\,\gamma^7) 
\end{bmatrix}	
\]
\[
 e^{\gamma^8 {\rm ad}_{A_8}} =
\begin{bmatrix}
1 & 0 & 0 & 0 & 0 & 0 & 0 & 0 \cr 
\sin^2 (\gamma^8) & \cos ( 2\,\gamma^8) & 0 & 0 & 0 & 
 0 & -\sin (2\,\gamma^8) & 0 \cr 
0 & 0 & \cos (\gamma^8) & 0 & 0 & \sin (\gamma^8) & 0 & 0 \cr 
0 & 0 & 0 & \cos (\gamma^8) & -\sin (\gamma^8) & 0 & 0 & 0 \cr 
0 & 0 & 0 & \sin (\gamma^8) & \cos  (\gamma^8) & 0 & 0 & 0 \cr 
0 & 0 & -\sin (\gamma^8) & 0 & 0 & \cos (\gamma^8) & 0 & 0 \cr 
-\cos (\gamma^8)\,\sin (\gamma^8)   & \sin (2\,\gamma^8) & 
 0 & 0 & 0 & 0 & \cos ( 2\,\gamma^8) & 0 \cr 0 & 0 & 0 & 0 & 0 & 0 & 0 & 1
\end{bmatrix}	
\]

Finally, the expression for the Wei-Norman formula is:
\[
\Xi = \begin{bmatrix}
1 & 0 & 0 & \sin ( 2 \gamma^3) & 0 & \xi_{16} & \xi_{17} & \xi_{18} \\
0 & 1 & 0 & 0 & 0 & \sqrt{\frac{2}{3}} \sin( \sqrt{6} \gamma^5 ) & 0 & \xi_{28} \\
0 & 0 & \cos( 2 \gamma^1 - \gamma^2 ) & -\sin( 2 \gamma^1 -\gamma^2) \cos ( 2 \gamma^3)  & 0 & \xi_{36} & \xi_{37} & \xi_{38} \\
0 & 0 & \sin( 2 \gamma^1 - \gamma^2 ) &  \cos( 2 \gamma^1 -\gamma^2) \cos ( 2 \gamma^3) & 0 &  \xi_{46} & \xi_{47} & \xi_{48} \\
0 & 0 & 0 & 0 & \xi_{55}  &  \xi_{56} & \xi_{57} & \xi_{58} \\
0 & 0 & 0 & 0 & \xi_{65}  &  \xi_{66} & \xi_{67} & \xi_{68} \\
0 & 0 & 0 & 0 & \xi_{75}  &  \xi_{76} & \xi_{77} & \xi_{78} \\
0 & 0 & 0 & 0 & \xi_{85}  &  \xi_{86} & \xi_{87} & \xi_{88} \\
\end{bmatrix}
\]
where the rather cumbersome explicit expression of various terms is included below for sake of completeness.
\begin{eqnarray*}
\xi_{55} & = &
\cos(2 \gamma^1 + \gamma^2) \cos(\gamma^3) \cos(\gamma^4) - \sin(2 \gamma^1 + \gamma^2) \sin(\gamma^3) \sin(\gamma^4) \\
\xi_{65} & = & 
\cos(\gamma^3) \cos( \gamma^4) \sin (2 \gamma^1 + \gamma^2) + \cos (2 \gamma^1 + \gamma^2) \sin (\gamma^3) \sin (\gamma^4) \\
\xi_{75} & = & 
-\cos (\gamma^1 - 2 \gamma^2) \cos (\gamma^4) \sin (\gamma^3) + \cos (\gamma^3) \sin (\gamma^1 - 2 \gamma^2) \sin (\gamma^4) \\
\xi_{85} & = & 
\cos (\gamma^4) \sin (\gamma^1 - 2 \gamma^2) \sin (\gamma^3) + \cos (\gamma^1 - 2 \gamma^2) \cos (\gamma^3) \sin (\gamma^4)
\end{eqnarray*}
\begin{eqnarray*}
\xi_{16} & = &
\frac{ \sin (\sqrt{6} \gamma^5) }{2 \sqrt{6} }
\left(2 + \cos (2 (\gamma^3 - \gamma^4)) + \cos (2 (\gamma^3 + \gamma^4))\right)\\
\xi_{36} & = &
\frac{\sin (\sqrt{6} \gamma^5) }{4 \sqrt{6}}
\left( -\cos (2 \gamma^1 - \gamma^2 + 2 \gamma^3 - 2 \gamma^4) + 
        \cos (2 \gamma^1 - \gamma^2 - 2 \gamma^3 + 2 \gamma^4) + 
        \cos (2 \gamma^1 - \gamma^2 - 2 (\gamma^3 + \gamma^4)) - \right. \\
&& - \left. 
        \cos (2 \gamma^1 - \gamma^2 + 2 (\gamma^3 + \gamma^4)) + 
        4 \cos (2 \gamma^1 - \gamma^2) \sin (2 \gamma^4)\right) \\
\xi_{46} & = &
\frac{\sin (\sqrt{6} \gamma^5)}{2 \sqrt{6} }
\left(\cos (2 \gamma^1 - \gamma^2 - 2 \gamma^4) - \cos (2 \gamma^1 - \gamma^2 + 2 \gamma^4) - 
        2 \cos (2 \gamma^1 - \gamma^2) \cos (2 \gamma^4) \sin (2 \gamma^3)\right)  \\
\xi_{56} & = &
-\cos (\sqrt{6} \gamma^5) (\cos (\gamma^3) \cos (\gamma^4) \sin (2 \gamma^1 + \gamma^2) + 
        \cos (2 \gamma^1 + \gamma^2) \sin (\gamma^3) \sin (\gamma^4)) \\
\xi_{56} & = &
\cos (\sqrt{6} \gamma^5)(\cos (2 \gamma^1 + \gamma^2) \cos (\gamma^3) \cos (\gamma^4) - \sin (2 \gamma^1 + \gamma^2) \sin (\gamma^3) \sin (\gamma^4)) \\
\xi_{76} & = &
-\cos (\sqrt{6} \gamma^5)(\cos (\gamma^4) \sin (\gamma^1 - 2 \gamma^2) \sin (\gamma^3) + \cos (\gamma^1 - 2 \gamma^2) \cos (\gamma^3) \sin (\gamma^4)) \\
\xi_{86} & = &
\cos (\sqrt{6} \gamma^5)(-\cos (\gamma^1 - 2 \gamma^2) \cos (\gamma^4) \sin (\gamma^3) + \cos (\gamma^3) \sin (\gamma^1 - 2 \gamma^2) \sin (\gamma^4)) 
\end{eqnarray*}
\begin{eqnarray*}
\xi_{17} & = &
\cos (2 \gamma^3) \cos (\gamma^6) \sin (2 \gamma^4) \sin (\gamma^5) - \cos (\gamma^5) \sin (2 \gamma^3) \sin (\gamma^6) \\
\xi_{37} & = &
-\cos (2 \gamma^1 - \gamma^2) \cos (2 \gamma^4) \cos (\gamma^6) \sin (\gamma^5) + 
  \sin (2 \gamma^1 - \gamma^2)\cdot \\ && \cdot
 \left(\cos (\gamma^6) \sin (2 \gamma^3) \sin (2 \gamma^4) \sin (\gamma^5) +  
        \cos (2 \gamma^3) \cos (\gamma^5) \sin (\gamma^6)\right) \\
\xi_{47} & = &
-\cos (2 \gamma^4) \cos (\gamma^6) \sin (2 \gamma^1 - \gamma^2) \sin (\gamma^5) - 
  \cos (2 \gamma^1 - \gamma^2)\cdot \\ && \cdot \left(\cos (\gamma^6) \sin (2 \gamma^3) \sin (2 \gamma^4) \sin (\gamma^5) + 
        \cos (2 \gamma^3) \cos (\gamma^5) \sin (\gamma^6)\right) \\
\xi_{57} & = &
-\sin (2 \gamma^1 + \gamma^2) \left(\cos (\gamma^3) \cos (\gamma^5) \cos (\gamma^6) \sin (\gamma^4) + 
        \cos (\gamma^4) \sin (\gamma^3) \sin (\gamma^5) \sin (\gamma^6)\right) + \\
&& + 
  \cos (2 \gamma^1 + \gamma^2) \left(\cos (\gamma^4) \cos (\gamma^5) \cos (\gamma^6) \sin (\gamma^3) - 
        \cos (\gamma^3) \sin (\gamma^4) \sin (\gamma^5) \sin (\gamma^6)\right)  \\
\xi_{67} & = &
\cos (\gamma^4) \sin (\gamma^3) \left(\cos (\gamma^5) \cos (\gamma^6) \sin (2 \gamma^1 + \gamma^2) + \cos (2 \gamma^1 + \gamma^2) \sin (\gamma^5) \sin (\gamma^6)\right) + \\
&& + 
  \cos (\gamma^3) \sin (
      \gamma^4) \left(\cos (2 \gamma^1 + \gamma^2) \cos (\gamma^5) \cos (\gamma^6) - \sin (2 \gamma^1 + \gamma^2) \sin (\gamma^5) \sin (\gamma^6)\right) \\
\xi_{77} & = &
\sin (\gamma^1 - 2 \gamma^2) \left(-\cos (\gamma^5) \cos (\gamma^6) \sin (\gamma^3) \sin (\gamma^4) + 
        \cos (\gamma^3) \cos (\gamma^4) \sin (\gamma^5) \sin (\gamma^6)\right) + \\
&& + 
  \cos (\gamma^1 - 2 \gamma^2) \left(\cos (\gamma^3) \cos (\gamma^4) \cos (\gamma^5) \cos (\gamma^6) + 
        \sin (\gamma^3) \sin (\gamma^4) \sin (\gamma^5) \sin (\gamma^6)\right) \\
\xi_{87} & = &
\cos (\gamma^3) \cos (
      \gamma^4) \left(-\cos (\gamma^5) \cos (\gamma^6) \sin (\gamma^1 - 2 \gamma^2) + \cos (\gamma^1 - 2 \gamma^2) \sin (\gamma^5) \sin (\gamma^6)\right) -\\
&& -
   \sin (\gamma^3) \sin (
      \gamma^4) \left(\cos (\gamma^1 - 2 \gamma^2) \cos (\gamma^5) \cos (\gamma^6) + \sin (\gamma^1 - 2 \gamma^2) \sin (\gamma^5) \sin (\gamma^6)\right)
\end{eqnarray*}
\begin{eqnarray*}
\xi_{81} & = &
 \cos(2 \gamma^7)
\left(\cos(\gamma^6) \sin(2 \gamma^3) \sin(\gamma^5) + \cos(2 \gamma^3) \cos(\gamma^5) \sin(2 \gamma^4) \sin(\gamma^6)\right) +  \\
&& + \frac{1}{3} 
 \left(\left( 2 + \cos(\sqrt{6} \gamma^5) \cos(\sqrt{6} \gamma^6)\right) \sin^2(\gamma^3)  + \right. \\
&& \left. + 
      \cos(2 \gamma^3)\left(\cos^2(\gamma^4) \left(-1 + \cos(\sqrt{6} \gamma^5) \cos(\sqrt{6} \gamma^6)\right) + 3 \sin^2(\gamma^4)\right)\right)\sin(2 \gamma^7) 
\\
\xi_{82} & = &
\frac{\sin(2 \gamma^7)}{3} \left(2 + \cos(\sqrt{6} \gamma^5)\ \cos(\sqrt{6} \gamma^6)\right) 
\\
\xi_{83} & = &
-\cos(2 \gamma^7) \left(\cos(2 \gamma^1 - \gamma^2) \cos(2 \gamma^4) \cos(\gamma^5) \sin(\gamma^6) + \right. \\ 
&& \left. +  
      \sin(2 \gamma^1 - \gamma^2)\left(\cos(2 \gamma^3) \cos(\gamma^6) \sin(\gamma^5)  -
  \cos(\gamma^5) \sin(2 \gamma^3) \sin(2 \gamma^4) \sin(\gamma^6)\right)\right) \\
&& 
+ \frac{\sin(2\ \gamma^7)}{6} \left(-4 + 
        \cos(\sqrt{6} \gamma^5) \cos(\sqrt{6} \gamma^6)\right) \cdot  
\\
&& \cdot \left(\cos(2\ \gamma^4)\ \sin(2\ \gamma^1 - \gamma^2)\ \sin(2\ \gamma^3) + \cos(2\ \gamma^1 - \gamma^2)\ \sin(2\ \gamma^4)\right) 
\\
\xi_{84} & = &
\cos(2 \gamma^7) \left(-\cos(2 \gamma^4) \cos(\gamma^5) \sin(2 \gamma^1 - \gamma^2) \sin(\gamma^6) + \right. \\ 
&& \left. +  
      \cos(2 \gamma^1 - \gamma^2) \left(\cos(2 \gamma^3) \cos(\gamma^6) \sin(\gamma^5) - 
            \cos(\gamma^5) \sin(2 \gamma^3) \sin(2 \gamma^4) \sin(\gamma^6)\right)\right) \\
&& + \frac{ \sin(2 \gamma^7)}{6} 
\left(-4 + \cos(\sqrt{6} \gamma^5) \cos(\sqrt{6} \gamma^6)\right) \left(-\cos(2 \gamma^1 - \gamma^2) \cos(2 \gamma^4) \sin(2 \gamma^3) + \sin(2 \gamma^1 - \gamma^2) \sin(2 \gamma^4)\right)
\\
\xi_{85} & = &
\cos(2 \gamma^7) \left(-\cos(\gamma^4) \sin(
          \gamma^3) \left(\cos(\gamma^5) \cos(\gamma^6) \sin(2 \gamma^1 + \gamma^2) + 
            \cos(2 \gamma^1 + \gamma^2) \sin(\gamma^5) \sin(\gamma^6)\right) +\right. \\ 
&& \left. +   
      \cos(\gamma^3) \sin(\gamma^4) \left(-\cos(2 \gamma^1 + \gamma^2) \cos(\gamma^5) \cos(\gamma^6) + 
            \sin(2 \gamma^1 + \gamma^2) \sin(\gamma^5) \sin(\gamma^6)\right)\right) + \\
&& + \frac{\sin(2 \gamma^7)}
{\sqrt{6} }\left(\cos(\sqrt{6} \gamma^6) \left(\cos(\gamma^3) \cos(\gamma^4) \sin(2 \gamma^1 + \gamma^2) + 
            \cos(2 \gamma^1 + \gamma^2) \sin(\gamma^3) \sin(\gamma^4)\right) \sin(\sqrt{6} \gamma^5) 
+ \right. \\ 
&& \left. +  \left(\cos(2 \gamma^1 + \gamma^2) \cos(\gamma^3) \cos(\gamma^4) - 
            \sin(2 \gamma^1 + \gamma^2) \sin(\gamma^3) \sin(\gamma^4)\right) \sin(\sqrt{6} \gamma^6)\right)
\\
\xi_{86} & = &
\cos(2 \gamma^7) \left(-\sin(2 \gamma^1 + \gamma^2) \left(\cos(\gamma^3) \cos(\gamma^5) \cos(\gamma^6) \sin(\gamma^4) + 
          \cos(\gamma^4) \sin(\gamma^3) \sin(\gamma^5) \sin(\gamma^6)\right) 
+\right. \\ 
&& \left. +   
    \cos(2 \gamma^1 + \gamma^2) \left(\cos(\gamma^4) \cos(\gamma^5) \cos(\gamma^6) \sin(\gamma^3) - 
          \cos(\gamma^3) \sin(\gamma^4) \sin(\gamma^5) \sin(\gamma^6)\right)\right) + \\
&& + \frac{\sin(2 \gamma^7)}
{\sqrt{6} } \left(\cos(\sqrt{6} \gamma^6) \left(-\cos(2 \gamma^1 + \gamma^2) \cos(\gamma^3) \cos(\gamma^4) + 
            \sin(2 \gamma^1 + \gamma^2) \sin(\gamma^3) \sin(\gamma^4)\right) \sin(\sqrt{6} \gamma^5) 
        + \right. \\ 
&& \left. +  \left(\cos(\gamma^3) \cos(\gamma^4) \sin(2 \gamma^1 + \gamma^2) + 
            \cos(2 \gamma^1 + \gamma^2) \sin(\gamma^3) \sin(\gamma^4)\right) \sin(\sqrt{6} \gamma^6)\right)
\end{eqnarray*}
\begin{eqnarray*}
\xi_{87} & = &
\cos(2 \gamma^7) \left(\cos(\gamma^3) \cos(\gamma^4)\left(\cos(\gamma^5) \cos(\gamma^6) \sin(\gamma^1 - 2 \gamma^2) - 
                \cos(\gamma^1 - 2 \gamma^2) \sin(\gamma^5) \sin(\gamma^6)\right) + \right. \\ 
&& \left. +  
          \sin(\gamma^3) \sin(\gamma^4)\left(\cos(\gamma^1 - 2 \gamma^2) \cos(\gamma^5) \cos(\gamma^6) + 
         \sin(\gamma^1 - 2 \gamma^2) \sin(\gamma^5) \sin(\gamma^6)\right)\right) \\
&& + \frac{\sin(2 \gamma^7)}
{\sqrt{6} } \left(\cos(\sqrt{6} \gamma^6) \left(\cos(\gamma^4) \sin(\gamma^1 - 2 \gamma^2) \sin(\gamma^3) + 
            \cos(\gamma^1 - 2 \gamma^2) \cos(\gamma^3) \sin(\gamma^4)\right) \sin(\sqrt{6} \gamma^5) + \right. \\ 
&& \left. +  \left(-\cos(\gamma^1 - 2 \gamma^2) \cos(\gamma^4) \sin(\gamma^3) + \cos(\gamma^3) \sin(\gamma^1 - 2 \gamma^2) \sin(\gamma^4)\right) \sin(\sqrt{6} \gamma^6)\right)
\\
\xi_{88} & = &
\cos(2 \gamma^7) \left(\sin(\gamma^1 - 2 \gamma^2) \left(-\cos(\gamma^5) \cos(\gamma^6) \sin(\gamma^3) \sin(\gamma^4) + 
            \cos(\gamma^3) \cos(\gamma^4) \sin(\gamma^5) \sin(\gamma^6)\right) + \right. \\ 
&& \left. +  
      \cos(\gamma^1 - 2 \gamma^2) \left(\cos(\gamma^3) \cos(\gamma^4) \cos(\gamma^5) \cos(\gamma^6) + 
            \sin(\gamma^3) \sin(\gamma^4) \sin(\gamma^5) \sin(\gamma^6)\right)\right) \\
&& + \frac{\sin(2 \gamma^7)}
{\sqrt{6} }
\left(\cos(\sqrt{6} \gamma^6) \left(\cos(\gamma^1 - 2 \gamma^2) \cos(\gamma^4) \sin(\gamma^3) - 
            \cos(\gamma^3) \sin(\gamma^1 - 2 \gamma^2) \sin(\gamma^4)\right) \sin(\sqrt{6} \gamma^5) 
    + \right. \\ 
&& \left. +  \left(\cos(\gamma^4) \sin(\gamma^1 - 2 \gamma^2) \sin(\gamma^3) + 
            \cos(\gamma^1 - 2 \gamma^2) \cos(\gamma^3) \sin(\gamma^4)\right) \sin(\sqrt{6} \gamma^6)\right) 
\end{eqnarray*}

This time the singularities of the Wei-Norman formula are obtained from the zeros of the following function:
\[
\det \Xi (\gamma) = \frac{1}{4} \cos(
      2 \gamma_3)  \cos(\sqrt{6} \gamma_5) \cos (2 \gamma_7 )   \left(2 + \cos( 2(\gamma_5 - \gamma_6) ) + \cos(2 (\gamma_5 + \gamma_6) ) \right)
\]

\section{Conclusion}
To be able to describe a time varying dynamics in terms of simple unitary operations is an important issue in quantum mechanics and it is foreseen that it will be a crucial one in quantum computation.
The method we propose here relates the one parameter flow of the Schr{\"{o}}dinger equation with an arbitrary decomposition of $SU(N) $ by computing the Jacobian of the coordinate transformation.
It is worth emphasizing that if the symbolic expression of the Wei-Norman formula rapidly explodes with the dimension of the system, its numerical integration can be easily and efficiently handled.
Furthermore, because of the way it is formulated (each parameter in \eqref{eq:Magnus-expan1} and \eqref{eq:WeiNorm-expan1} lives on $\mathbb{R}^1 $ or on $ \mathbb{S}^1$), the numerical integration will give a solution which respects the group structure, regardless even of the rounding error. 
Finally, notice that, as a byproduct, the explicit expressions for the exponentials of the adjoint operators should be useful in the study of the solution of the density operator equation, which is often expanded in terms of unitary superoperators.

 \bibliographystyle{abbrv}
\small


\end{document}